# Comparison of grain texture in round Bi2212 and flat Bi2223 superconducting wires and its relation to high critical current densities

F. Kametani, J. Jiang, M. Matras, D. Abraimov, E. E. Hellstrom and D. C. Larbalestier

*Applied Superconductivity Center, National High Magnetic Field Laboratory, Florida State University, Tallahassee FL 32310 USA*

**Abstract**

Why $Bi_2Sr_2CaCu_2O_x$ (Bi2212) allows high critical current density $J_c$ in round wires rather than only in the anisotropic tape form demanded by all other high temperature superconductors is important for future magnet applications.  Here we compare the local texture of state-of-the-art Bi2212 and Bi2223 $((Bi,Pb)_2Sr_2Ca_2Cu_3O_{10})$, finding that round wire Bi2212 generates a dominant *a*-axis growth texture that also enforces a local biaxial texture (FWHM <15°) while simultaneously allowing the *c*-axes of its polycrystals to rotate azimuthally along and about the filament axis so as to generate macroscopically isotropic behavior.  By contrast Bi2223 shows only a uniaxial (FWHM <15°) *c*-axis texture perpendicular to the tape plane without any in-plane texture.  Consistent with these observations, a marked, field-increasing, field-decreasing $J_c(H)$ hysteresis characteristic of weak-linked systems appears in Bi2223 but is absent in Bi2212 round wire.  Growth-induced texture on cooling from the melt step of the Bi2212 $J_c$ optimization process appears to be the key step in generating this highly desirable microstructure.

**Introduction to**

Superconducting magnets require high critical current density $J_c$ in long-length wire forms, which are inevitably polycrystalline [1]. One of the most significant barriers to applying high temperature superconducting (HTS) materials in random polycrystalline forms has been due to the current-obstructing effects of randomly oriented grain boundaries (GBs) [1,2]. Thus the two HTS conductors available in the market today, $(Bi, Pb)_2Sr_2Ca_2Cu_3O_x$ (Bi2223) and $YBa_2Cu_3O_{7-\delta}$ (YBCO) (or more generally REBCO, where RE denotes rare earth) are fabricated in complex ways determined by the need to minimize grain misalignments so as to reduce or even eliminate high angle GBs (HAGBs). High $J_c$ in Bi2223, which allows operation at 77 K due to its high critical temperature $T_c$ of ~110K [3], requires a complex deformation- and reaction-induced texturing process which aims to preferentially align the c-axes of grains normal to the tape plane [4]. Present Bi2223 tapes have a uniaxial texture of ~15° full-width at half-maximum (FWHM), allowing a $J_c$ of order 500 A mm$^{-2}$ at 77 K and self-field [5,6,7]. REBCO coated conductors [8,9,10,11] have taken much attention away from Bi2223 because their growth on quasi-single crystal templates allows a much better, biaxial texture (2-5° FWHM) which generates much higher $J_c$ (77 K, sf) of ~$3 \times 10^4$ A mm$^{-2}$ [12]. However, a major impediment to magnet applications of these HTS conductors is their 20-40:1 shape anisotropy, their large electromagnetic anisotropy and their delivery in fixed sizes determined by their complex manufacturing process, rather than by the desires of their users. On the other hand, $Bi_2Sr_2CaCu_2O_x$ (Bi2212) is the only HTS cuprate in which high $J_c$ can be developed in a round wire form [13, 14] which also allows it to be supplied in multiple multifilament architectures [15] and in the twisted state that benefits low hysteretic losses, isotropic properties and high magnetic field quality. Another huge advantage of round wires is that they can also be flexibly cabled into conductors of arbitrary current capacity, as is well demonstrated by many uses of Nb-Ti and $Nb_3Sn$ wires [16]. The opportunity to develop similar capabilities in a round wire cuprate conductor with high current density capability to fields well above 50 T, more than twice that possible with any Nb-based conductor, suggests the value of a detailed understanding of how Bi2212 differs from its anisotropic HTS sibling Bi2223. Although its $T_c$ of 90-95 K restricts applications at 77 K, its high irreversibility field $H_{irr}$ below 10 K [17] and its round wire form are exactly the desirable characteristics required for high field magnets. Especially because $J_c$ of Bi2212 has now reached 3200 and 2500 A mm$^{-2}$ at 12 and 20 T, 4.2 K in fully dense, round wire form with an irreversibility field $H_{Irr}$(4K) > 100 T [18,19,20], it has become a very serious candidate for high field NMR [21,22] and particle accelerators such as future upgrades of the Large Hadron Collider (LHC) [23].

The Bi2212 round wires have long been assumed to be macroscopically untextured. Indeed, since high $J_c$ could only be developed in very short lengths, it never received much attention and many assumed that the low $J_c$ values of long length wires were due to its roundness which was presumed to deny the macroscopic texture believed to be essential to high $J_c$. In fact, the key breakthrough to high $J_c$ in Bi2212 round wires was to understand that the principal critical current-limiting mechanism (CLM) was blockage of supercurrent flow by filament-diameter bubbles formed during the melt step of heat treatment [18,19,20,24,25,26], and not by HAGBs generated by any lack of macroscopic texture, as seemed totally plausible considering the need to make Bi2223 and REBCO in tape forms so as to generate strong crystallographic texture. Considering that supercurrent flow occurs preferentially in the $CuO_2$ planes of all HTS cuprate materials, Bi2212 included, we felt that it was time to open the issue of Bi2212 texture in round wires again. Actually we know that planar, [001] tilt thin film bicrystal GBs of YBCO, Bi2212 and Bi2223 all possess very similar exponential decay of intergranular $J_c$ as a function of misorientation angle [2]. Whether such experiments are valid for much more general GBs is less clear. For example, subsequent work has shown that non-planar GBs in REBCO have much higher $J_c$ than their planar GB counterparts [27, 28] but the effect of the non-planarity is only to about double the critical angle for current blocking from about 3° to 6-7° rather than make a transformational change in GB behavior. Nevertheless, this relaxation of critical angle is what enables *ex situ* coated conductor processes to generate high $J_c$ [29]. Thus the question of whether a similar process was at work in filaments of Bi2212 presented itself to us as an urgent question now that we have been able to establish a reliable overpressure process (OP) for making very high $J_c$ Bi2212 round wires [18]. We wanted to understand the true grain structure of these OP round wires so as to get a more general understanding of the role of naturally occurring GBs formed by the controlled solidification from the melt originally developed by Heine *et al.* in 1989 [30]. OP does densify the Bi2212 filaments (thus making preparation of samples much easier for the EBSD examination that is central to this paper), making their physical connectivity high, but OP does not change the filament shape [18,19,20] and it seems hard to imagine that their grain structure is enhanced by the densification.

The motivation for this paper is therefore to address a long-standing conundrum: why is high $J_c$ possible in round wire Bi2212 but not in round wires of its close sibling Bi2223 [31]. To perform the study, we used Electron Backscatter Diffraction Orientation Imaging Microscopy (EBSD-OIM) to reveal the texture of state-of-the-art high $J_c$ Bi2223 tape and overpressure processed round wire Bi2212. We found that while the Bi2223 tape shows the expected uniaxial texture of about 15° around [001], the Bi2212 round wire is actually not macroscopically untextured. In fact it possesses a strong *a*-axis texture parallel to the filament axis that imposes thus too a marked biaxial texture with similar 15° in-plane spread (the *b* direction [010] is thus normal to the wire axis). Electromagnetic isotropy is enforced by a gradual rotation of the *c*-axis of grains around the azimuth about the wire axis such that a conductor containing hundreds of filaments is locally anisotropic, while isotropic as a whole. This unique grain structure, especially the in-plane biaxial texture, correlates well to the much higher $J_c$ of Bi2212 compared to Bi2223, as well as to the lack of weak-link signature in the $J_c(H)$ characteristics *of* Bi2212 and its presence in Bi2223 that we here present as a striking difference between the two conductors. The study also opens up the intriguing possibility of making isotropic, multifilamentary round wires of other HTS materials if similar grain and GB structures could be realized in them too. This would be especially powerful for compounds with the REBCO structure because of their inherently much lower electronic anisotropy and higher irreversibility field, properties that would allow operation in high fields, even at liquid nitrogen temperatures.

**Results**

In order to clarify the often complex and sometimes confusing ways of referring to texture in these wires, we need a few words about the principal ways that we present our experimental data in this paper. The external perspective of the experimenter naturally defines two principal axes, the axis along the wire or that normal to the wire (or the tape). In crystallographic parlance, the wire axis view is normally called the rolling direction (RD), while the perpendicular or side view made at 90° to RD is called the normal direction (ND). In the case of the tape there are clearly two preferred NDs, one parallel to the broad face of the tape (as used here) and one perpendicular. For a round wire there is no preferred ND but, since we found that the *c*-axis rotates around the wire axis, it is possible to choose a local ND such that low index planes dominate in the ND view, as they do in either preferred-direction ND view of a Bi2223 tape. This is what we do here. Figure 1a further relates the macroscopic, wire-level view to the crystallographic view by showing the Bi2212 and Bi2223 crystal structures drawn with VESTA software [32]. Both are perovskites in which oxide layers of Bi, Sr, Ca and Cu stack alternately in an inherently anisotropic structure. We note that this produces an inherent electronic anisotropy in both Bi2212 and Bi2223 that can be hidden if there is a random grain orientation distribution. Accordingly, we use both in-plane and out-of-plane projections on inverse pole figures (IPF) to define the individual grain misorientations, as shown in figure 1b. For in-plane rotations or misorientations, the *a*- or *b*-axes of the crystal ([100] or [010]) change direction while keeping the direction of the *c*-axis [001] constant, while only out-of-plane rotations are needed to change the *c*-axis orientation, as is illustrated in figure 1c. As such, we can visualize both in-plane and out-of-plane components of the grain misorientations as we show later in Figures 3-8. It should be noted that the a and b lattice parameters are so close that [100] cannot be distinguished from [010] by EBSD. However, Bi2212 is known to form plate-like grains with a high *a/b* aspect ratio, presumably because of the lattice modulation along the b-axis [33, 34]. Thus, in this paper, we presume that Bi2212 grains are always longer along the a-axes than the b-axes. In this interpretation the dominant texture of Bi2212 is then [100] along the wire axis (RD), while [010] lies normal to the wire axis.

We first wished to compare the $J_c$ behavior of Bi2212 and Bi2223, as shown in Figure 2. Two striking features are clear: one is that the $J_c(H)$ at 4.2 K of round wire Bi2212 is about 3 times higher than that of Bi2223 and second that Bi2223 flat tape showed a significant dependence of the magnitude of $J_c(H)$ according to whether it was measured in increasing or decreasing field. As usual, $J_c(H)$ was always higher in decreasing field (figure 2 inset). Strikingly, since such hysteresis is normally attributed to the presence of weak links in the polycrystalline network [35], the Bi2212 round wire showed no $J_c(H)$ hysteresis (figure 2b). Although both conductors are fully (>95%) dense because OP densification was used in both [5, 6, 7, 18], it is clear that weak links are present in Bi2223, but also that the signature is apparently absent in Bi2212.

Figure 3a and c compare longitudinal SEM cross sections of the Bi2223 flat filament viewed parallel to the broad filament face (ND, as illustrated) with that of a Bi2212 filament also viewed along ND. Grey regions of Bi2223 or Bi2212 are dominant in both cases with occasional darker alkaline earth cuprate (AEC) or lighter, low-$T_c$ phase (Bi2212 and Bi2201 for Bi2223 and Bi2212, respectively) regions in the cross-sections. The Bi2223 filament is remarkably phase-pure, whereas the Bi2212 filament is less phase-pure, having larger AEC, Cu-free (CF) or Bi2201 2$^{nd}$ phase regions, as well as a small residual void fraction. The grain textures and grain-to-grain misorientations of the same regions are visualized as inverse pole figures (IPF) and GB maps in Figures 4-6. In figure 3b and d, the Bi2223 and Bi2212 grains are colored based on their grain orientation when viewed parallel to ND. Due to the strong uniaxial *c*-axis texture of the Bi2223 tape, the plate-like Bi2223 grains appear as thin laths with ab-planes parallel to ND. Their grains are typically ~0.3-1 μm thick and 5-20 μm in diameter. Because of their marked uniaxial texture, almost all Bi2223 grains appear green, blue or their mixture, indicating a rather continuous in-plane misorientation distribution between [100] and [110] about a common [001] axis. The dominant GBs are straight and generally parallel to the ab-plane of either of the adjacent grains and to the tape surface. Indeed, GBs parallel to the c-axis are very rare. The uniaxial [001] texture that is the object of the deformation- and growth-induced texture processing route is quite clear.

As mentioned above, in Bi2212 there is no preferred axis of ND, except that it is normal to the wire axis. Here we chose a length of filament where there was a dominant set of Bi-2212 grains with *ab*-planes lying parallel to ND, for which the ND-IPF map of the Bi2212 filament (figure 3d) shows a dominant green color corresponding to a dominant [010] orientation. As for Bi2223, the Bi2212 grains are also plate-like in shape and of similar thickness to Bi2223, but they are ~50-300 μm in length, almost 10 times larger than the Bi2223 grains. Note however that the lateral growth of the grains is strongly constrained by the nominal 15 μm diameter filaments. Even though some grain growth from filament to filament occurs, few grains are wider than 30 μm. As with Bi2223, the dominant GBs are those lying parallel to the ab-planes. As for Bi2223 and also for Bi2212, the grains have a strong tendency to stack on top of each other in a colony structure in which all grains share a common *c*-axis. We made a 600 μm long ND-IPF cross section montage of this Bi2212 filament and confirmed that the GBs //c-axis are very rare (later in Figure 8 we will show one such cluster of misoriented grains discovered in this scan).

A way to compare the Bi2212 and Bi2223 texture relevant to their grain boundary connectivity is by coloring GBs according to their local grain-to-grain misorientation. In this case we choose a misorientation of 20° (figure 4a and b). Figure 4b makes it immediately clear that the majority of Bi2212 GBs have local misorientations ≤20° (green), while GBs with >20° misorientation (dark brown) are significantly more evident in Bi2223 (figure 4a). A further comparative view of the texture can be obtained from binning the individual grain-to-grain misorientations, as shown in Figure 4c and d, which plot the misorientations of all grains examined in the EBSD scans. In the Bi2223 flat filament, the grain-to-grain misorientation angles, although peaking at ~12°, are broadly distributed from <5° to 45° (figure 4c). On the other hand, the distribution of Bi2212 grain-to-grain misorientation angles shows a sharp peak around 8° (figure 4d). We note that figure 4c and d contain angles >45°, which occurs when the out-of-plane texture component dominates the misorientation, as in some Bi2212 colony GBs. In Bi2223, due to the uniaxial texture, the dominant component of misorientations is in-plane, so the angular range is confined to 45° because of the inability of the EBSD to distinguish between [100] and [010] due the very small difference of the *a* and *b* lattice parameters in either compound.

The most significant way of comparing the difference between the Bi2212 and Bi2223 grain textures may be however by their ND and RD inverse pole figures. As already shown in the ND-IPF grain map of figure 3, the distribution of Bi2223 grain orientations is *ab*-plane dominant in both the ND and RD projections. Comparing figure 5a and b, the uniaxial texture viewed along the RD which perceives rotations normal to the RD is slightly better than along the ND. The out-of-plane grain misorientations are <10° along the filament, whereas this distribution creeps up to ~15° across the filament. However, both the ND- and RD-IPF share the same random in-plane distributions, as judged by the uniform distribution of orientations between [100] and [110] in figure 5a and b (recall that [100] and [010] cannot be distinguished by EBSD). This is a natural characteristic of a uniaxial texture in which only the c-axis of grains is aligned normal to the tape surface.

By contrast a rather different picture emerges from the inverse pole figures of Bi2212 in figure 5c and d. The two very distinctive features are a cluster of points within about 15° of [100] in two orthogonal directions in the RD view but a rather continuous distribution of grain orientation points in the ND view within 15° of the axis connecting [100] and [001]. What this latter distribution signifies is that there is a continuous rotation of the [001] axis of Bi2212 grains along the wire axis. Although no positional information is present in figure 5, it is clear from extended scans (like the 600 μm long one mentioned earlier) that there is a continuous rotation of [001] along the filament axis as new Bi2212 grains grow on solidifying from the melt. Grains cluster within ~15° of [100] in the in the *ab*-plane. Figure 5d shows that the grain orientations along the wire axis (RD) are also preferentially [100] with only a ~15° variation of misorientation in both in- and out-of-plane directions and this makes clear that the Bi2212 grains also possess a biaxial texture, the *a*-axis ([100]) lying parallel to RD which is also the filament axis. As noted above, the [001] *c*-axis can rotate easily about the filament axis, allowing a continuous range of orientations perpendicular to the wire direction. Another feature of the Bi2212 grain structure is plastic bending and/or twisting of individual grains. The ND-IPF of figure 5c shows that many grains have orientations that continuously vary between the [100]/[110] and [001] directions due to this grain plasticity that must occur during grain formation and growth at high temperature.

**Discussion**

The key microstructural finding of this paper is that there is a very marked texture in melt-processed Bi2212 round wire and that the texture is, rather surprisingly, biaxial. From a superconducting property point of view, the $J_c$(H,4 K) of the Bi2212 conductor is about 3 times higher than that of the Bi2223 tape conductor, in spite of the significantly higher $T_c$ of Bi2223 (110 versus ~80 K) . As has been reported before [36], there is a markedly hysteretic $J_c(H)$ characteristic for this state-of-the-art Bi2223 conductor, but surprisingly we find almost no weak link signature in the Bi2212 conductor.

A key benchmark for applications is that overall conductor current density $J_E$ must exceed 300-500 A/mm$^2$ in a domain of H and T of practical interest and it is clear that both Bi2223 and Bi2212 can do this at 4K to fields of well over 20 T, given that superconductor filling fractions are of order 40% for Bi2223 and 25-30% for Bi2212. But, from a future development point of view, it would be highly desirable to understand what fraction of the superconductor cross-section actually carries current. Since both conductors have been heat treated under an external overpressure, they are both well over 95% dense, thus ruling out voids as current-limiting obstacles in either conductor and most plausibly implicating GBs as the dominant current-limiting mechanism. However, it is notoriously hard to evaluate the extent to which the $J_c$ developed by flux pinning within grains is throttled by blocking grain boundaries, non-superconducting second phases, cracks or other obstacles. The striking differences in texture shown by this study suggest a first focus on the distribution of misorientations within the grain boundary network.

There are two principal and different models that have been proposed to describe supercurrent flow in uniaxially textured Bi2223 flat tapes, the Brick-wall [37] and Railway-switch model [38]. In the Brick-wall model, the $J_c$ of the current path is limited by *c*-axis current flow across basal-plane faced, *c*-axis twist GBs, resulting in a Josephson-like weak link behavior that can however sustain finite $J_c$ in high fields because of the very large surface area of the basal-plane–faced GB junctions [37]. On the other hand, the railway-switch model postulates that the many small-angle, c-axis tilt GBs found in Bi2223 where *a*- or *b*-axes of grains terminate at the broad *ab*-plane face of an adjacent grain provide a strongly coupled, vortex-pinning dominated current path [38]. Neither mechanism is believed to allow supercurrent flow over 100% of the cross-section and there is great uncertainty on what the active cross-section might be, because measurements, either by transport or by magnetization, measure the product of $J_c$ and A (where A is the actual section carrying supercurrent, rather than the total measured cross-section) without either $J_c$ or A being known independently. Bulaevskii *et al*. have laid out procedures for separating the different current paths [39], but a significant problem is that current transfer across the matrix Ag can confuse the shapes of the V-I curves that are an important part of the evaluation procedure. Earlier we did succeed in measuring V-I curves on an extracted, bare Bi2223 filament with about half the $J_c$ of the present conductor and did conclude that they gave evidence of *c*-axis transport [40]. More generally, however, Bulaevskii *et al.* concluded that vortex pinning dominates the V-I curves and $J_c(H)$ characteristics rather than the Josephson-like Brick-wall behavior [39]. In any case it seems most likely that both mechanisms operate and that a complex series-parallel current path determines the actual superconducting characteristics. The weak-link hysteresis in Bi2223 observed here supports such an interpretation (note that this hysteresis was much larger in earlier conductors of lower overall $J_c$ [36]). The very intriguing possibility raised by the quasi-biaxial texture of the round Bi2212 wires which show almost no hysteresis is that their higher $J_c$ can indeed be explained by, not just correlated to, their better texture.

Within the framework of the exponential decay of $J_c$ with misorientation angle θ [2] in planar bicrystals of all cuprates, it is quite natural to believe that the much better, quasi-biaxial texture of Bi2212 compared to Bi2223 shown in Figure 5 would correlate well to the lack of weak-link behavior in Bi2212 and its three times higher $J_c$ (figure 2). Figure 6 shows that that magnetization $J_c(H)$ behavior of both Bi2212 and Bi2223 is rather well linearized by the Kramer function ($\Delta m^{1/2}(\mu_0 H)^{1/4}$, where m and H are the magnetic moment and field, respectively) used to describe vortex depinning by flux line lattice shear [41], which again is suggestive of a dominant, strongly coupled current path. All of these data are consistent with a larger fractional cross-section current path in 2212 and fewer weak-linked grain boundaries in the biaxially aligned Bi2212 filament.

Under this interpretation, the reason why the two textures are so different becomes of prime importance. One key difference between the two compounds is in their processing. In the case of Bi2223, the powder is a mixture of Bi2212 and "stuff', mostly Ca and Cu oxides. Formation of Bi2223 occurs largely by solid state diffusion (though a small amount of liquid is present) after a complex, rolling-induced alignment of the Bi2212 grains in the precursor, which then convert *in situ* to Bi2223 during heat treatments. Bi2223 is often described as possessing a deformation *and* reaction-induced texture. In Bi2223 tapes, the filament width is ~150-250 μm [7], far larger than the diameter of Bi2223 grains (5-20 μm, as is seen in figure 3b and 4a), thus allowing freedom for the *a*- or *b*-axis of grains to grow in any direction within the *ab*-plane without degrading the uniaxial rolling texture. By contrast, the Bi2212 texture develops on slow cooling after melting of the starting Bi2212 powder. Large Bi2212 grains grow up to several 100 μm long along the ab-plane, a size ~10 times larger than the filament diameter which is typically 15-20 μm. Perhaps crucially, Bi2212 grains tend to grow faster along the *a*- than the *b*-axis, and much faster than along the *c*-axis, thus forming very thin grains with a high in-plane *a/b* aspect ratio > 5, rather than the round disks of Bi2223 [34]. But growth from the melt confers another benefit, because the surrounding Ag restricts the longest grains to those with the fastest growth *a*-axes aligned along the filament axis. This preferred growth within the narrow filament cavities results in both in-and out-of-plane grain orientation along the wire axis, as is seen in figure 5d. The almost round cross section of filament cavities also allows the grains to grow with their *c*-axis oriented in any direction normal to the wire axis (ND). As the Bi2212 grains grow, the multiple regions that have the local biaxial texture can form in parallel. They can join with each other having different azimuthal orientations, as is shown by the reddish regions in figure 7b. However, the ND-IPF of figure 5c and d indicate that the plastic bending and twisting of Bi2212 grains that appear on the ND-IPF as continuously changing orientation plots can presumably compensate large misorientations over long lengths by local, small misorientations. In addition, our earlier TEM studies suggested that the Bi2212 ab-planes can easily bend in distances of ~10-20 nm so as to minimize the out-of-plane misorientations.

It should be noted that Bi2212 round wires are not as perfectly biaxial as the REBCO coated conductors in which the out-of-plane and in-plane misorientation is minimized to less than 5° [12]. Indeed, as is clearly seen in figure 4, the general misorientation of the GBs in Bi2212 is typically 10-15° which is significantly higher than the critical angle (2-3°) found in experiments on thin film bi-crystals with planar GBs parallel to the film normal [2]. But, in spite of this higher misorientation, the weak link signature of Bi2223 is absent in Bi2212 (and in biaxially textured coated conductors too), implying that Bi2212 GBs are less obstructive. Three additional factors may play an important role in enhancing current flow across Bi2212 round wire GBs. One is the very large, basal-plane-faced GB area per grain. As an earlier study of YBCO coated conductors pointed out, the $J_c$ reduction across GBs becomes less pronounced as the GB area increases due to the larger current-carrying cross section [27]. Actually we estimate the area of basal-plane-faced GBs in the Bi2212 to be ~500-5000 $\mu m^2$ per grain due to the large *ab*-plane grain boundary area. This is very favorable compared to the transverse cross-section of ~200 $\mu m^2$ for the whole filament. The second factor is that the GBs in these Bi2212 round wires are almost parallel to the direction of macroscopic current flow. Because of the rotation of the *c*-axis about the wire axis, there will always be regions where any *c*-axis current across these basal-plane GBs will be parallel to the magnetic field in a Lorentz force-free configuration, which minimizes depinning of weakly pinned vortex segment at GBs [27,42]. The third factor is the possibility to significantly overdope Bi2212 well beyond that possible in Bi2223 or YBCO, which is certainly beneficial for increasing the superfluid density at the GB and for strengthening the vortex line tension, both of which enhance the GB current density [43]. We thus conclude that multiple factors enhance the capability for developing higher $J_c$ in Bi2212 round wires that operate much less effectively in Bi2223 tapes. A vital one is almost certainly the favoring of *a*-axis growth along the filament axis from which the biaxial texture evolves.

**Conclusion**

In summary, we have presented the grain textures of state-of-the-art Bi2223 tapes and Bi2212 round wires. In contrast to the uniaxially textured Bi2223 tape in which the in-plane GB misorientations are essentially random with an out-of-plane misorientation or *c*-axis texture of <15°, the Bi2212 round wires exhibit a quasi-biaxial texture as a result of the growth of long, high aspect ratio grains within the narrow filament cavities. Along the wire axis, the in-plane GB misorientation in Bi2212 is constrained within ~ 15°, and to <15° for the out-of-plane misorientation too. In addition, the large grain size allows large-area, *ab*-plane basal-plane-faced GBs favorable for intergrain current transport. Our study demonstrated that a unique biaxially aligned microstructure is present in high $J_c$ Bi2212 round wire which suggests the possibility of making round wires from other HTS materials if a similar melt-driven path can be found for them too.

**Methods**

The Bi2212 wire of 0.8 mm diameter was fabricated by Oxford Superconducting Technology using the Powder-in-Tube (PIT) technique. A typical conductor (though many architectures are possible [15]) uses an 85 filament first stage drawn into hexagonal form, 18 of them being then stacked inside a Ag alloy tube to make a final size conductor containing ~1500 (85x18) filaments, each about 15 μm in diameter. For our EBSD study, we used a specially designed 27x7 variant of this architecture described in our previous paper [24]. The Bi2212 wires were heat treated by the OP technique [18]. The Bi2223 tape was fabricated by Sumitomo Electric Inc. also using a PIT process. In contrast to the Bi2212 wire fabrication, rolling of the Bi2223 wire was performed both before the first heat treatment and between the first and second heat treatment in order to optimize the deformation- and reaction-induced uniaxial texture [6]. The final tape was ~4.3 mm wide by 0.17 mm thick. $I_c$ values were measured at 4.2 K applying magnetic field up to 15 T by the four probe method. The $I_c$ values are converted to $J_c$ using the stated filling factor of 40% for the Bi2223 tape and that measured on cross-sections of the Bi2212 wires densified by OP just before melting when the filaments are still round. The $J_c$ data presented here are actually for a Bi2212 round wire of 85x17 architecture which was heat-treated by the same OP technique. This wire had a $J_c(H)$ characteristic within 10% of that exhibited by earlier measurements of the 27x7 architecture wire.

Longitudinal cross sections were prepared by mechanical grinding and polishing on SiC papers and diamond lapping film, a final gentle polish with 50 nm colloidal alumina, followed by ion-milling at 2.0 kV in a Gatan PECS ion mill. Electron Backscatter Diffraction Orientation Imaging Microscopy (EBSD-OIM) to visualize the filament grain structure was performed in a Carl Zeiss 1540EsB scanning electron microscope with an EDAX Hikari camera, and TSL OIM Collection software.

**Acknowledgments**

This work was supported by the US Department of Energy (DOE) Office of High Energy Physics under grant number DE-SC0010421, by the National High Magnetic Field Laboratory (which is supported by the National Science Foundation under NSF/DMR-1157490), and by the State of Florida. We acknowledge the help of V. S. Griffin in Vibrating Sample Magnetometer measurements and N. C. Craig for transport critical current measurements. We are also grateful for discussions with U. P. Trociewitz at NHMFL, M. Rikel at NEXANS and C. Scheuerlein at CERN.

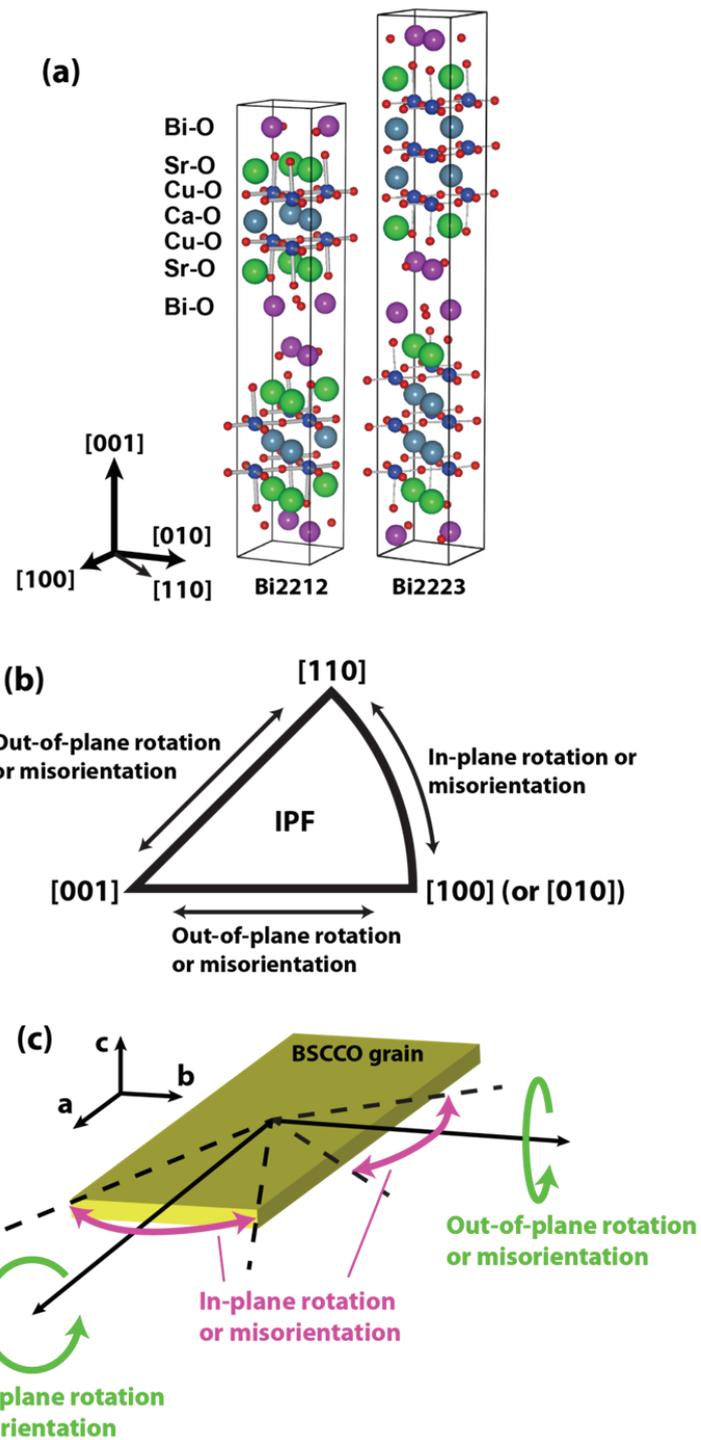

Figure 1 (a) Crystal structures of Bi2212 and Bi2223 and definition of unit axes. (b) Location of grain misorientations on the inverse pole figure (IPF). (c) Schematic illustration of in-plane and out-of-plane rotations of the BSCCO grains. The broad flat surface of a BSCCO grain is always parallel to its ab-plane. Note that the axes of out-of-plane rotations can take any directions containing its ab-plane.

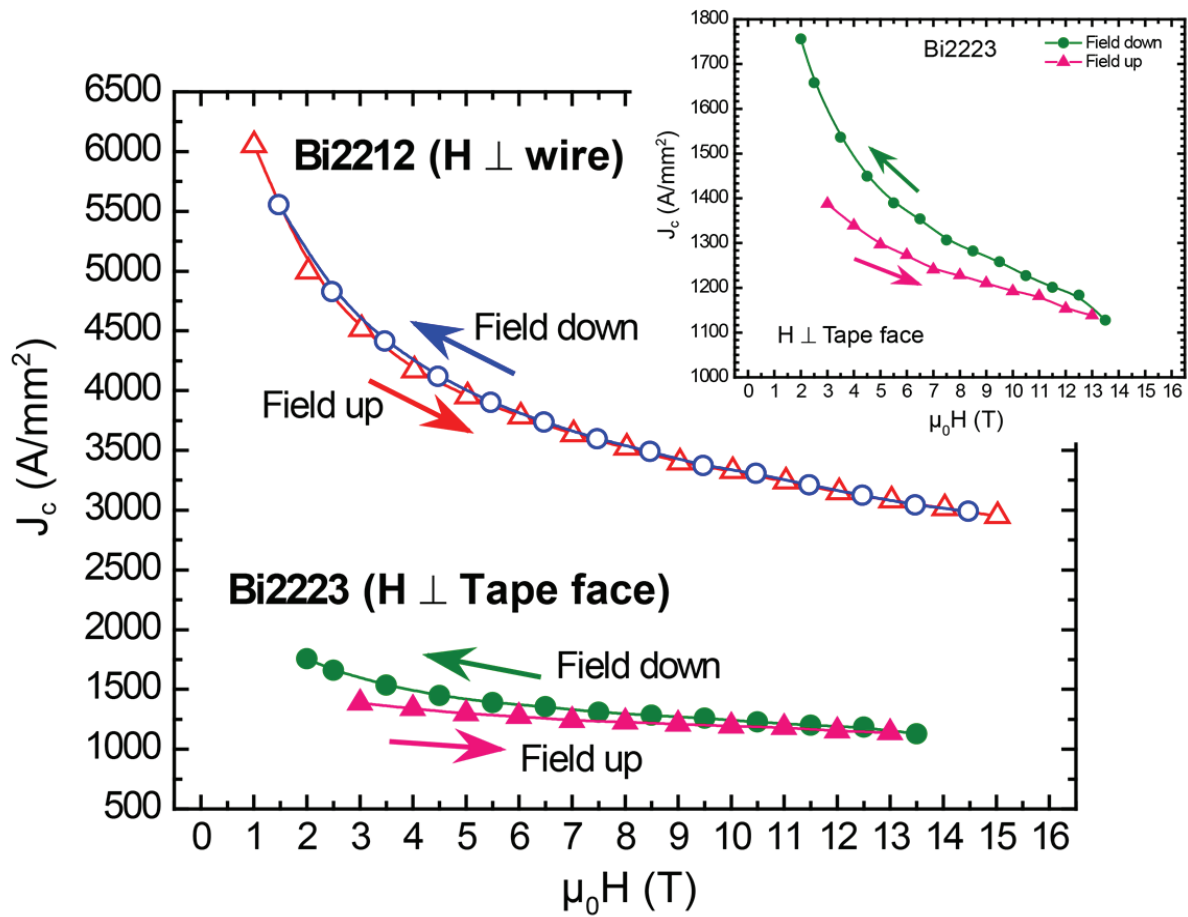

Figure 2 Comparison of the superconducting critical current density $J_c$(H, 4.2K) of the Bi2223 flat tape and the Bi2212 round wire. The field H was applied normal to the wire axis for Bi2212, and to the tape face for Bi2223, respectively. The red and pink symbols of Bi2212 and Bi2223 indicate $J_c$ measured in increasing field, whereas the blue and green symbols of Bi2212 and Bi2223 represent $J_c$ in decreasing field. Note that $J_c$ of Bi2212 round wire is approximately 3 times larger than that of the Bi2223 flat tape and is without any field hysteresis, which is quite evident in Bi2223 and is shown more clearly in the inset.

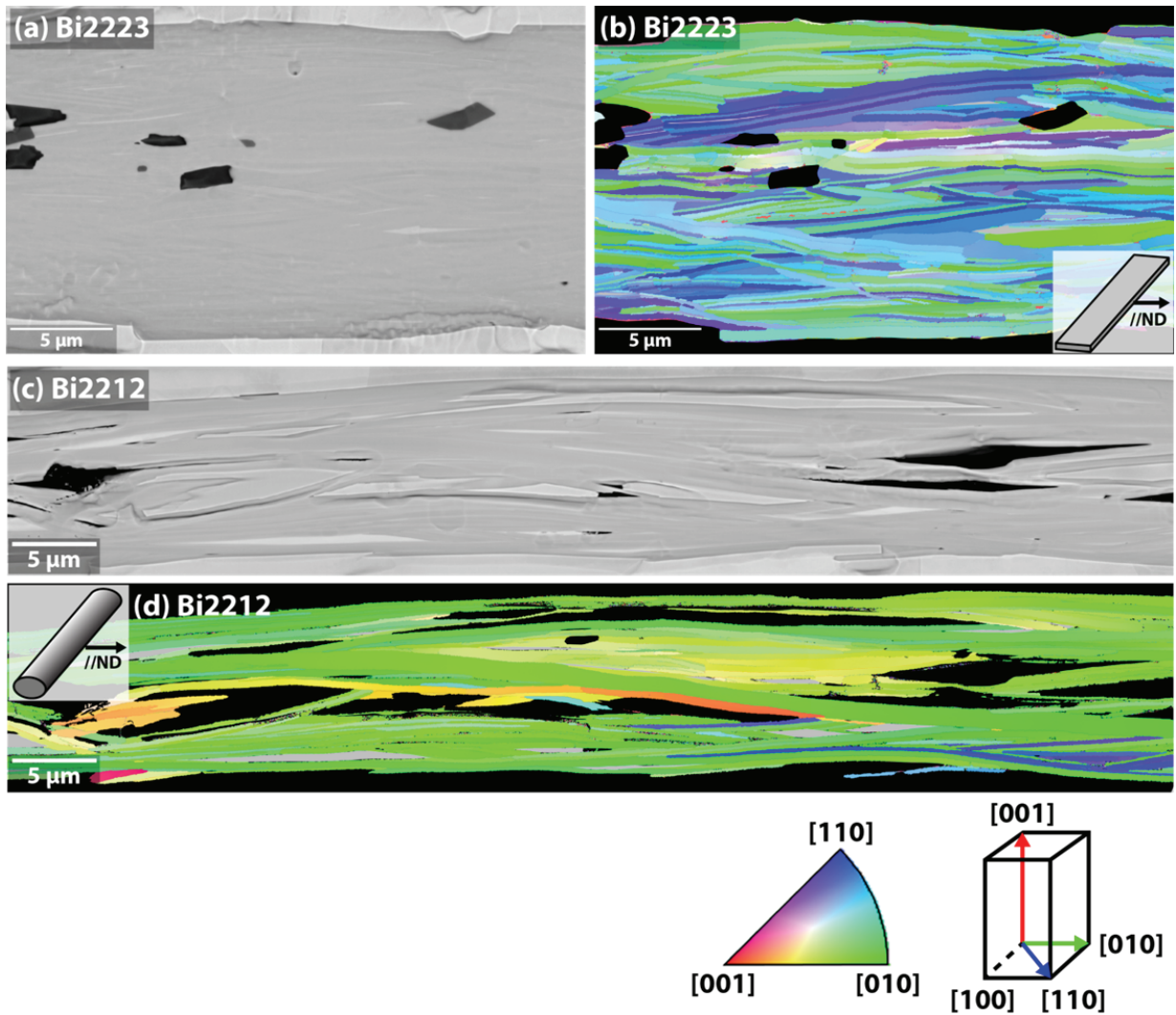

Figure 3 Comparison of the grain structure of Bi2223 and Bi2212 filaments as viewed along the ND parallel to the tape plane in Bi2223 and along an *ab*-plane-oriented section of a Bi2212 filament. A longitudinal cross section of a representative area of Bi2223 is shown in (a) by a backscattered Electron SEM image, and in (b) by a corresponding ND-IPF map, whereas those of Bi2212 are shown in (c) and (d). In the ND-IPF maps, Ag and second phase regions are blacked out and low $T_c$ phase regions (Bi2212 in Bi2223 and Bi2201 in Bi2212) are colored gray. Colors in (b) and (d) correspond to the grain orientations defined by the crystallographic directions parallel to the ND projection axis which is perpendicular to the wire direction.

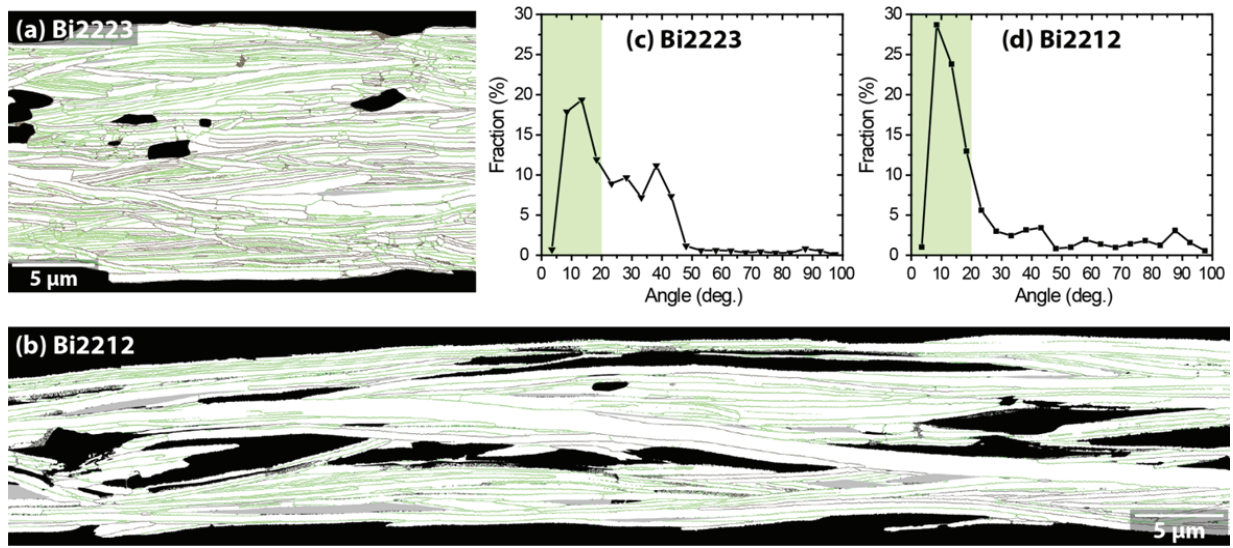

Figure 4 (a) (b) GB maps of the areas of figure 3a (Bi2223) and 3c (Bi2212), respectively. GBs with misorientation angle of ≤20° are traced in green, while misorientations >20° are colored in dark brown. Based on these GB maps, the fractional GB length is plotted as a function of misorientation angle for (c) Bi2223 and (d) Bi2212. The shaded area represents the fraction of GBs having misorientations of ≤20° that appear green in (a) or (b). Note the large difference in the most frequent misorientation (~14° in Bi2223 versus ~8° in Bi2212) between the two conductors.

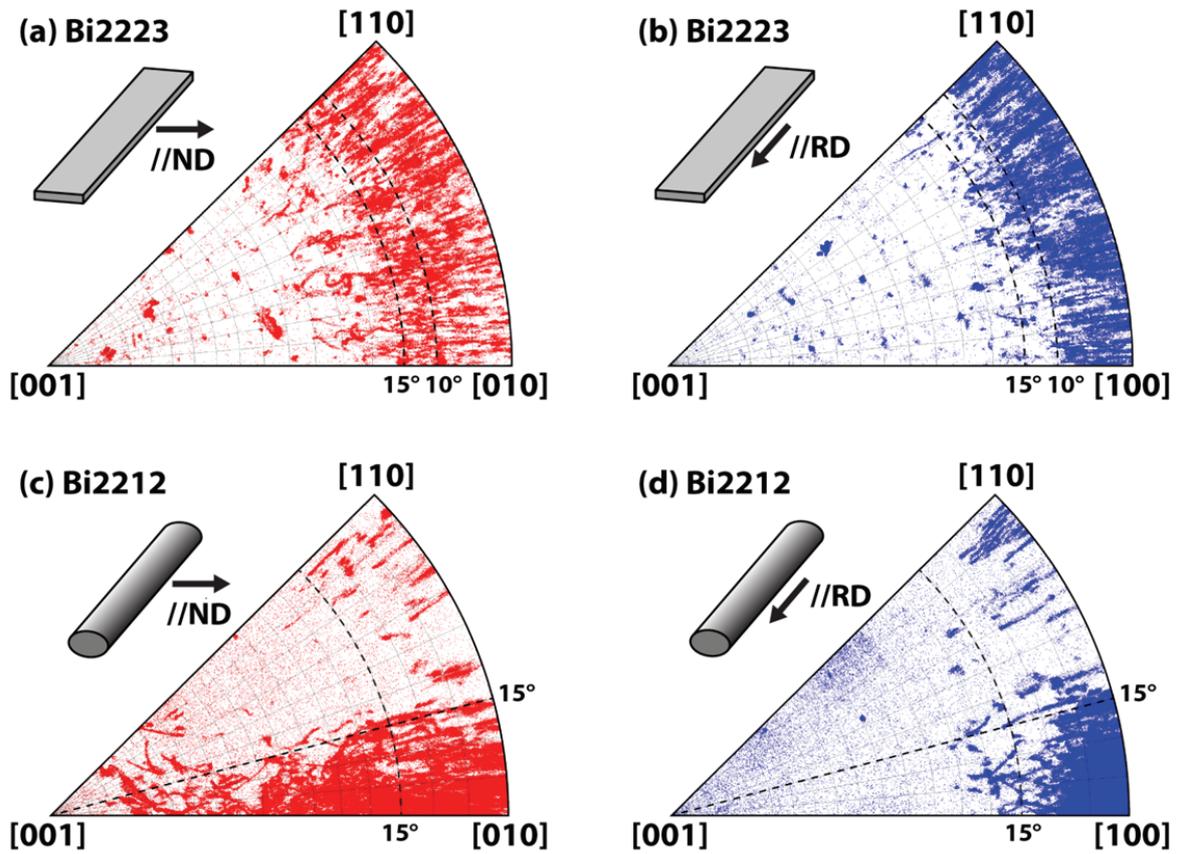

Figure 5. ND and RD Inverse Pole Figures (IPF) of the grain orientations in Bi2223 and Bi2212 are shown in (a)(b) and (c)(d), respectively. The IPFs of Bi2223 are derived from the image of figure 3b, whereas those of Bi2212 are a combination of figure 3d and 7b. They are stereographic projections of the grain orientations parallel to the ND projection axis (which is normal to the filament direction and parallel to the tape plane for Bi2223) in (a) and (c), and in (b) and (d) the RD which lies parallel to the filament direction. The black dotted lines in (a) and (b) mark the dominant misalignments of 15° and 10° away from the ab-plane that defines the dominant [001] texture of Bi2223, while the very different 15° "corner-pocket" texture of Bi2212 that defines a significant biaxial, in-plane alignment around [100] is shown in (d).

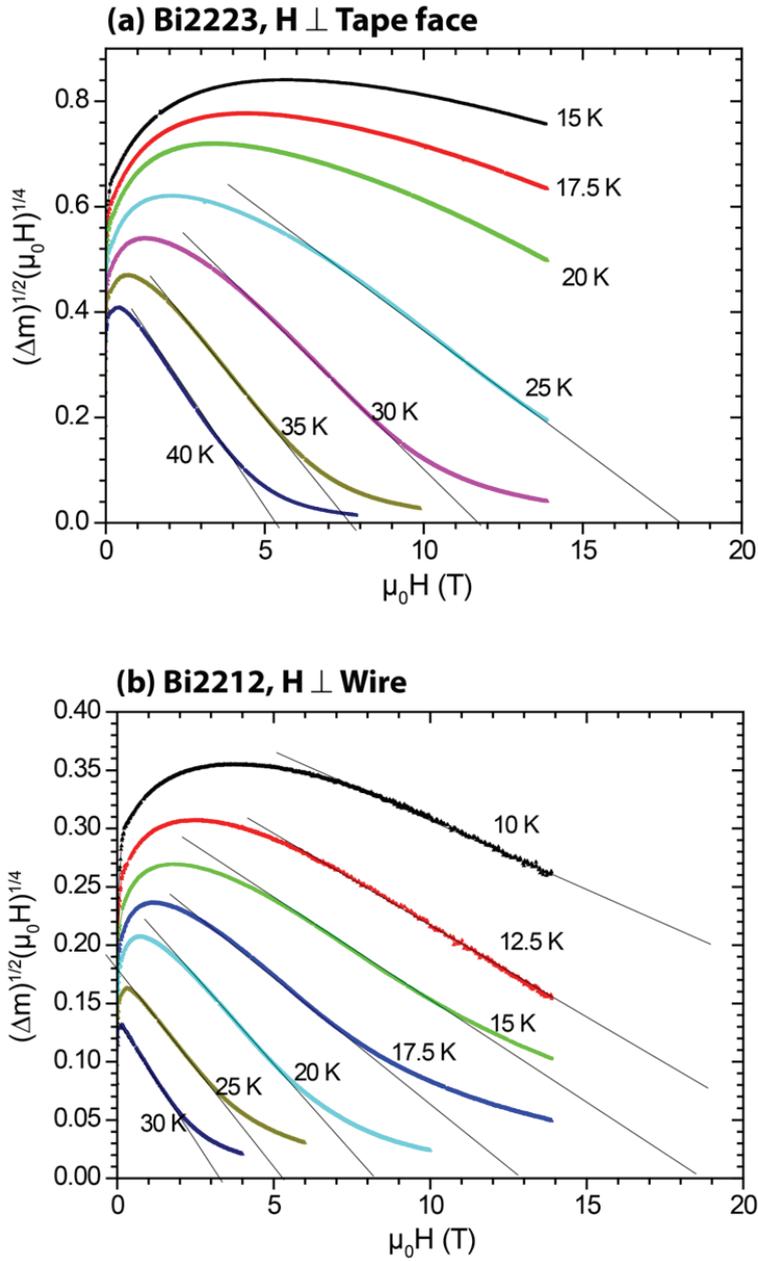

Figure 6 Kramer function plots of the Bi2212 round wire and Bi2223 flat tape, derived from magnetization measurements over the temperature range 10-30 K for Bi2212 in (a) and 15-40 K for Bi2223 in (b). A considerable linear regime exists for both conductors, which is generally taken to be consistent with a vortex pinning dominated $J_c$ behavior. Δm = m(field down) – m(field up), where m denotes the magnetic moment. Note that the magnetization $J_c$ is proportional to Δm.

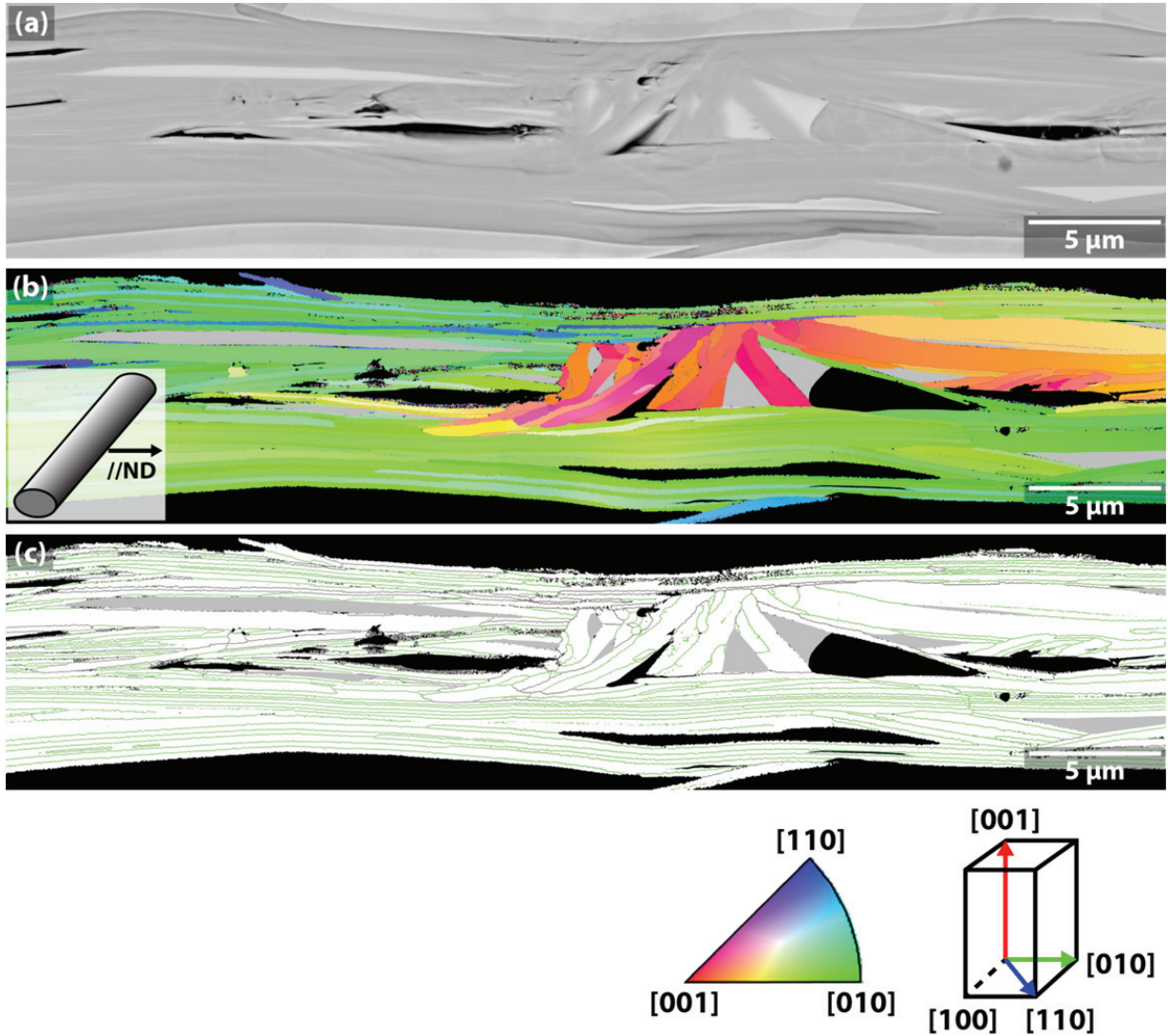

Figure 7 SEM image, ND-IPF and GB maps of another portion of a 600 μm long longitudinal cross section, a portion of which was also shown in Figure 3c and d, showing that comparatively rare, but highly misaligned regions do exist in the Bi2212 filament. As shown in (b), most grains are well textured and in this case green, which denotes that [010] is parallel to the ND. There is also a highly misoriented red-orange region indicating that c-axis [001] oriented grains are present. (c) Although the majority of GBs are colored green (<20°), the colony boundaries have much larger misorientation, signifying that they contain a large component of out-of-plane misorientation.